\documentclass[letter,twocolumn]{jpsj2} %% for letters

\def\<{\langle}
\def\>{\rangle}
\def\({\left(}
\def\){\right)}
\def\[{\left[}
\def\]{\right]}
\def\up{\uparrow}
\def\dn{\downarrow}
\def\s{\sigma}

\title{Pressure-Induced Phase Transition to a Novel Spin State in Striped Nickelates}
\author{Eiji Kaneshita$^1$ and Alan R. Bishop$^2$}
\inst{
$^1$Argonne National Laboratory, Argonne, IL 60439\\
$^2$Los Alamos National Laboratory, Los Alamos, NM 87545}

\abst{
We analyze pressure effects on stripe states within a selfconsistent Hartree-Fock calculation for a model of a striped nickelates.
The results show a transition induced by high pressure and predict possible new spin states.
We describe characteristics in the phonon excitations at the predicted transition, based on a real-space random phase approximation.
}
\recdate{\today}
\kword{
stripes,
nickelate,
pressure effect,
local mode,
phonon}

\begin{document}
\sloppy
\maketitle

%\section{Introduction}
Much recent attention has been devoted to stripe patterns~\cite{machida,kato,zaanen1,poiblanc,schulz}, which are a typical spin and charge ordering in many transition metal oxide compounds:  La$_{2-x}$Sr$_{x}$CuO$_4$~\cite{tranquada1,tranquada2}, La$_{2-x}$Sr$_{x}$NiO$_4$~\cite{sachan,tranquada3,shlee}, etc.
The stripe state has been intensely studied due to possible implications for high-temperature superconductivity.

To investigate the stripe state, many kinds of experiments have been carried out under a variety of conditions:  different hole dopings, several kinds of impurities, etc.
Neutron scattering is a powerful tool to detect stripes and such experiments have revealed important properties of the stripe state.
For example, hole-doping effects on stripes has been studied by neutron scattering, in terms of the doping dependence of the stripe periodicity and orientation~\cite{matsuda}.

In addition to the doping effects, pressure should yield important information on the stripe states.
However, a pressure study is not readily compatible with neutron scattering experiments.
Due to the difficulties of a high-pressure study on striped compounds, not many experiments have been performed.
Nevertheless, high-pressure effects open up possibilities for finding new features and for understanding the properties of stripe states.
Therefore, theoretical work is desirable to give guidance for high-pressure experiments.
Here, we provide a prediction of a new stripe state induced by high pressure based on a Hartree-Fock (HF) calculation for a model of striped nickelates with $\frac{1}{3}$-hole doping.

%\section{model}
We consider the hamiltonian
\begin{align}
\mathcal{H}=& \mathcal{H}_0
+ \mathcal{H}_{\mathrm{Ni}} + \mathcal{H}_{\mathrm{O}},\\
%%%
\mathcal{H}_0 =& \sum_{i,j,m,n,\s} \[t_{ij,mn}(u_k)
+ e_{i,m}\delta_{ij}\delta_{nm}\]\,c^\dagger_{im\s}c_{jn\s},\\
%%%
\mathcal{H}_{\mathrm{Ni}} =& \sum_{i,m}(U+2J)\, n_{im\up}n_{im\dn}\nonumber\\
&{}+ \sum_{i,\s} \[U\, n_{iA \s} n_{iB -\s}
+  (U-J)\, n_{iA\s} n_{iB\s}\]\nonumber\\
&{} + \sum_{i,m\neq n} c^\dagger_{im\up}c_{in\up}
\[J \, c^\dagger_{in\dn} c_{im\dn}
-   J'\, c^\dagger_{im\dn} c_{in\dn}\],\\
%%%
\mathcal{H}_{\mathrm{O}} =& \sum_{i,m}\frac{1}{2M}\,p_{im}^2+\sum_{i,m}\frac{K}{2}u_{im}^2,
\end{align}
where $i$ and $j$ indicate a unit comprised of a Ni and two O sites, and $m$ and $n$ denote the orbitals ($A \equiv d_{x^2-y^2}$, $B \equiv d_{3z^2-r^2}$, $p_{x}$ or $p_{y}$) or specify an O ion (O$_x$ or O$_y$) when we consider ionic motion.
Note that the summations in $\mathcal{H}_{\mathrm{Ni}}$ $(\mathcal{H}_{\mathrm{O}})$ are taken only on the Ni (O) sites.
The Ni on-site energy $e_{i}$ is equal to $-\Delta \pm E_z$, where $\Delta=\epsilon_p-\epsilon_d$ and $E_z$ is the energy splitting of $A$ and $B$ orbitals caused by the crystal field.
The term $J'$ is a pair hopping, and its energy $J'$ is set equal to the Hund exchange energy $J$ for simplicity.
$t_{ij,mn}(u_k)$ is non-zero only for nearest neighbor hoppings and depends on the  displacement of the involved oxygen-ion $u_k$ through the electron-phonon coupling constant $\alpha$:
$t_{ij,mn}(u_k)=t_{mn}-\alpha u_k$ for $t_{ij,mn}(u)\neq0$ cases.
For simplicity, we restrict the displacements of O ions only in the Ni-O bond direction.

We solve the model selfconsistently within a HF approximation in a supercell of size $6 \times 6$ with periodic boundary conditions.
To simulate the pressure change, we introduce a pressure parameter $\lambda$, which is set as $\lambda=1.0$ for ambient pressure and increased for higher pressure.
The hopping energy is varied through $\lambda$.
We take the following parameter set~\cite{yonemitsu,zaanen}: $t_{Ap}=\lambda$, $t_{Bp}= \pm\frac{\lambda}{\sqrt{3}}$,  $\Delta=9$, $Ez = 1$,  $J = J' =1$, and $U  = 4$ in energy unit $t_0$ eV; and $\alpha=4t_0^{\frac{3}{2}}$ eV/{\AA} and $K=32t_0^2$ eV/{\AA}$^2$.
We set the lattice constant to $a=1$, and all the energies are in units of $t_0$.
In real oxide materials~\cite{zaanen}, $t_0$ is in the range 1.3-1.5.

We also calculate the linear lattice dynamics from a renormalized electron Green's function within an adiabatic approximation~\cite{yonemitsu}.
From the eigenmodes of the lattice oscillation, we plot the corresponding dynamic neutron scattering cross section~\cite{martin,kane}:
\begin{align}
S(\mathbf{k},\omega)=\int dt\, e^{-i\omega t}
\sum_{l l^\prime}{\langle e^{-i \mathbf{k R}_l(0)}
e^{i\mathbf{k R}_{l'}(t)}\rangle}.
\label{eq:Skw}
\end{align}

%\section{results}
\begin{figure}[htbp]
\begin{center}
\includegraphics[width = 0.9\linewidth]{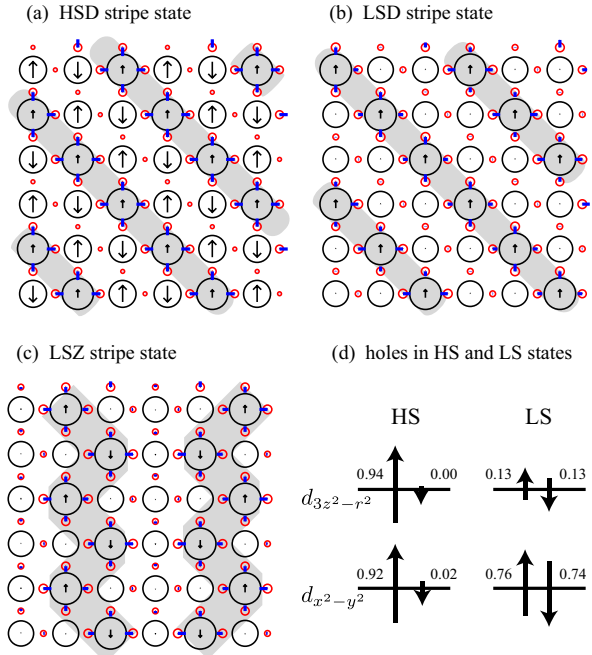}
\caption{(Color online) Different stripe states:  (a) HSD, (b) LSD, and (c) LSZ.
The circles (radius) and the arrows (length and direction) represent the hole and spin densities, respectively.
Lines at the positions of O-ions represent static displacements.
The shadows show the stripes.
These are plotted for pressure parameters in the range 1.33-1.35.
The different hole states in the AF (in the HSD state) and NM (in the LSZ state) domains are depicted in (d).
The numbers represent the up- or down-spin hole densities calculated with $\lambda=1.30$ for HSD and $\lambda=1.33$ for LSZ.
} \label{fig:gs}
\end{center}
\end{figure}
At ambient pressure ($\lambda=1.0$), the groundstate is calculated to be a diagonal stripe state, as experimentally observed~\cite{sachan,tranquada3,shlee}.
The spin and charge configurations of the diagonal stripe state are shown in Fig.~\ref{fig:gs}(a), where we observe the stripes with anti-ferromagnetic (AF) domains between them.
Here, we term the state with such AF domains a \textit{high-spin} (HS) stripe state in order to distinguish it from another stripe state which consists of almost non-magnetic (NM) domains [see Fig.~\ref{fig:gs} (b)].
We term the latter a \textit{low-spin} (LS) stripe state.
Details of these spin states will be discussed below.
The stripes we find have \textit{diagonal}, \textit{vertical}, and \textit{zigzag} characters.

\begin{figure}[htbp]
\begin{center}
\includegraphics[width = 0.8\linewidth]{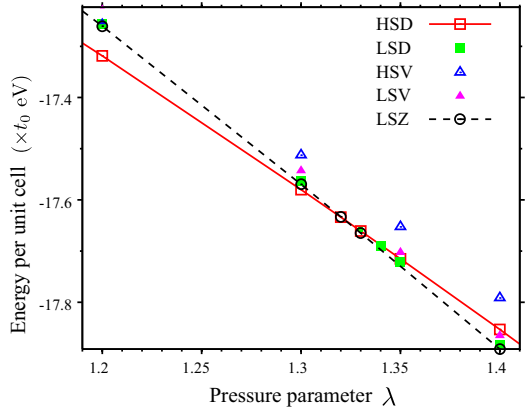}
\caption{(Color online) Energies of different stripe states for different $\lambda$.
The periodicity is 3 sites for all of the states:  one hole per stripe.
The lowest-energy state changes from HSD to LSZ at $\lambda\sim1.32$.
} \label{fig:energy}
\end{center}
\end{figure}

First, we show that the HS diagonal (HSD) and LS zigzag (LSZ) stripe states [Figs.~\ref{fig:gs}(a) and (c)] are HF groundstates in our model.
In Fig.~\ref{fig:energy}, the energies of these states are plotted for different $\lambda$, together with other stripe states: the LS diagonal (LSD), the HS vertical (HSV), and the LS vertical (LSV) stripe states.
The periodicity of the stripes is 3 sites for all states; i.e., there is one hole per stripe.
There is a transition from the HSD to the LSZ stripe state around $\lambda=1.32$.
The LSD state has a close but slightly larger energy than the LSZ state.

Next, we consider details of the transition.
We first focus on the change of the stripe orientation and discuss changes of the local spin state separately below.
In our previous study on a Cu$_2$O$_3$ ladder system~\cite{kane3}, we found the tendency that the zigzag configuration becomes preferable to the diagonal one under high pressure, although only the HS states appear in these cuprate systems.
Therefore, we suppose that the transition from LSZ to LSD would not occur even when the pressure is increased further.
In fact, the energy difference of these two states becomes larger as pressure increases and this is consistent with the behavior in the ladder system.
However, if a transition from HSD to LSD exists for some reason (e.g., different model parameters, different doping, incomensurability, etc.), we would find the transition from LSZ to LSD at a higher pressure.

Both the HSV and LSV stripe states cannot be the groundstate within the region we consider here.
However, the LSV stripe state might become more stable in the higher pressure region by analogy with the Cu$_2$O$_3$ ladder case~\cite{kane3}.
Therefore, we discuss the properties of the HSD, LSZ and LSD stripe states, including the possibility of the HSD-LSD transition.

There are distinct differences in the spin and hole densities between the HS and LS states.
Clear differences are found in the AF or NM domain between the stripes [See Fig.~\ref{fig:gs}(d)].
In the AF domain of the HSD state, one site, which has four local hole states (two $d$ orbitals with two spins), is nearly half-filled:  $\sim1.90$ holes ($\approx$ 48\% filling) at the site.
The hole density on such a site is equally distributed into the two orbitals ($d_{x^2-y^2}$ and $d_{3d^2-r^2}$).
Here the hole states in the different orbitals on the same site are dominated by one spin:  The dominant spin state has more than 97\% of the holes in the orbital.
The densities and ratio between the local hole states do not change much for $1.0\leq\lambda\leq1.32$

On the other hand, the NM sites in the LSZ state are occupied by $\sim1.80$ holes each ($\approx$ 45\% filling).
Also, there is no longer equal distribution into the two orbitals:  $\sim85$\% in $d_{x^2-y^2}$ and $\sim$15\% in $d_{3d^2-r^2}$.
The up- and down-spin states are almost equally occupied:  One of them has 55\% of the holes in the orbital.
The LSD state has a similar hole distribution to the LSZ.

As for the physical picture of the transition from HSD to LSD, it would be considered that the pressure enhances the hopping of holes and the system tends to be rather metallic.
This effect should prefer the NM domain in order to gain the kinetic energy.
The metal-insulator transition, however, cannot be treated in our model properly due to the small system size.

Thus far, we have shown that our calculation predicts a transition between the HS (HSD) and LS (LSZ) states.
Now we discuss how we can observe the transition, taking into account the possibility of the HSD-LSD transition.
X-ray diffraction should be able to detect a change in stripe orientation from diagonal to zigzag as a change of the charge order from $(\frac{2\pi}{3},\frac{2\pi}{3})$ to $(\frac{\pi}{3},\pi)$.

However, it would be difficult to observe the change of the local spin density itself.
Thus, if the transition does not change the stripe orientation (HSD-LSD), it is difficult to detect directly.
Therefore, we consider the observation of the transition from a different perspective, namely, lattice dynamics.
Theoretical~\cite{martin,kane2} and experimental~\cite{mcqueeney,reznik} studies show that stripe states are accompanied by characteristic phonons due to the inhomogeneous structure of charge and spin densities coupled to ion positions.

\begin{figure}[htbp]
\begin{center}
\includegraphics[width = 1.0\linewidth]{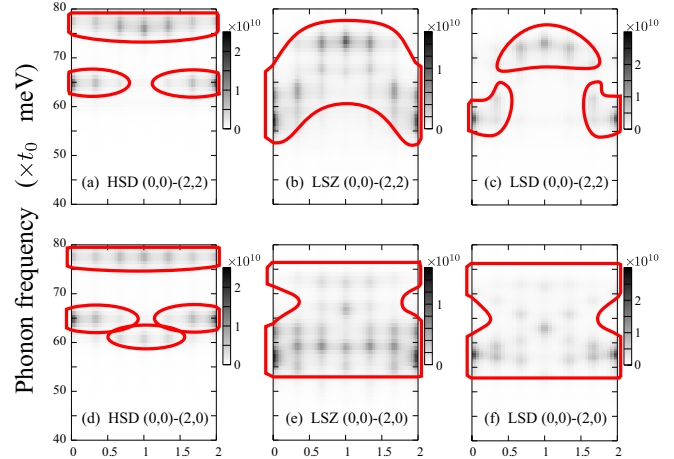}
\caption{(Color online) Phonon spectra for HSD, LSZ, and LSD stripe states.
Two different paths are plotted: (0,0)-(2,2) [(a), (b), (c)] and (0,0)-(2,0) [(d), (e), (f)], where $\pi=1$.
The results are symmetrized with respect to $x$ and $y$ directions. The pressure parameter is set as $\lambda=1.30$ for HSD [(a), (d)], $1.33$ for LSZ [(b), (e)], and $1.345$ for LSD stripes [(c), (f)].
The lines are guides to the eye.
} \label{fig:phonon}
\end{center}
\end{figure}

In Fig.~\ref{fig:phonon}, we plot the phonon spectra $S(\mathbf{k},\omega)/|\mathbf{k}|^2$ [Eq.~(\ref{eq:Skw})] of the HSD, LSZ, and LSD stripe states, calculated near the transition point.
In the LS cases, the whole spectra are softened compared to those of the HSD state.
The softening is $\sim7 t_0$ meV ($\approx 56t_0$ cm$^{-1}$) at $(0,0)$ and $\sim15 t_0$ meV ($\approx 121t_0$ cm$^{-1}$) at $(\pi,0)$ in the LSD case.
In the HSD case, phonon eigenmodes exist around $61t_0$, $65t_0$, and 75-78 $t_0$ meV near the transition point.
At $\lambda=1$ (data not shown in figure), the corresponding branches to those at $\sim 61t_0$ and 65 $t_0$ meV are at $\sim55t_0$ and $60t_0$ meV, respectively, and they are shifted upwards in frequency as $\lambda$ increases.
The 75-78 $t_0$ meV branch does not change much.

In the LSD case, the phonon eigenmodes lie in the range 45-78 $t_0$ meV.
However, most of the modes have much weaker intensity than the strongest mode, which we will discuss later:
In Fig.~\ref{fig:phonon}, we only see the spectra of high-intensity modes.

Along the path (0,0)-$(2\pi,2\pi)$, most of the intensity is around $65t_0$ and 75-78 $t_0$ meV for the HSD case, and around $56t_0$ and 70-73 $t_0$ meV for the LSD case.
Along the path (0,0)-$(2\pi,0)$, the HSD case has additional modes around $61t_0$ meV; and the LSD case has high intensity in the range 55-65 $t_0$ meV.

\begin{figure}[htbp]
\begin{center}
\includegraphics[width = 0.95\linewidth]{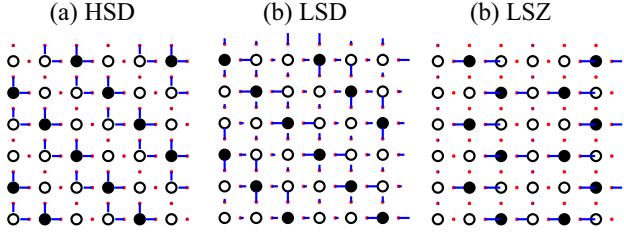}
\caption{(Color online) Lattice excitation pattern of high-intensity modes for (a) HSD case at $\sim$65$t_0$ meV, (b) LSD case at $\sim$58$t_0$ meV, and (c) LSZ case at $\sim$56$t_0$ meV.
The filled circles represent the stripe sites.
} \label{fig:mode1}
\end{center}
\end{figure}

The highest intensity modes for the HSD and LSD cases are at $65t_0$ and $58t_0$ meV, respectively.
Both modes are found at $(0,0)$ and show the same oscillation pattern [Figs.~\ref{fig:mode1}(a) and (b)].
These modes involve the O ions around the stripes and should be connected to the charge sliding of the diagonal stripes.
(A similar mode has been calculated for cuprates in Ref.~\cite{martin} --- these are distinct from the breathing-type~\cite{mcqueeney} modes.)

The highest intensity mode in the LSZ is found at $56t_0$ meV, and this is also related to the stripe sliding [Fig.~\ref{fig:mode1}(c)].
In the LSZ case, however, the intensity of this mode is not so high as in the other cases, and there are several modes with similar intensity.
For example, one of the modes involves all of the O$_y$ ions around the stripes, oscillating at $58t_0$ meV.
There are some modes involving off-stripe ions, as well.
We discuss the lower energy modes, other than the sliding one, below.

We have seen that the high-intensity modes are the same sliding-related oscillation in both the HSD and LSD cases, and also found a similar mode in the LSZ.
Here, we discuss the differences of the low-energy modes in the HS and LS cases, in the energy ranges 60-70 $t_0$ meV (HSD) and 55-65 $t_0$ meV (LSD, LSZ).

\begin{figure}[htbp]
\begin{center}
\includegraphics[width = 0.8\linewidth]{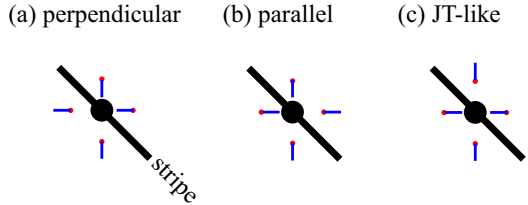}
\caption{(Color online) Lattice vibration pattern around stripes.
The thick line shows the stripe direction.
The vibration pattern of local modes in the HS case are represented by a combination of these three patterns:
(a) perpendicular to the stripe, (b) parallel to the stripe, (c) JT type.
} \label{fig:local}
\end{center}
\end{figure}

In the HS case, all of these modes are spatially localized and involve only the O ions around the stripes.
The vibration pattern of these modes is well-reproduced as a combination of three local vibrations (Fig.~\ref{fig:local}).
The local vibration patterns in Figs.~\ref{fig:local}(a) and (b) have the center-of-mass shift perpendicular and parallel to the stripes, respectively.
The one in Fig.~\ref{fig:local}(c) is a Jahn-Teller (JT)-like vibration.

In the LS cases, the modes in the 55-65 $t_0$ meV range include non-local modes as well, which involve all or most of the O ions.
The phonons for the LS cases do not have as much of the original stripe character as those for the HS case, although there are still such modes, including the sliding one.
It would be more accurate to say that the LS cases have some non-stripe modes in the lower-energy range, while the HS one has such modes but they are all in the higher-energy range, separated from the lower-energy stripe modes.
However, since the correspondence of the modes is weak between the HSD and LSD states (except the sliding one), it is difficult to definitively assign each mode.

From the analysis of the calculated phonon spectra, we conclude that the phonon excitations in the LS stripe state have less stripe signature than in the HS stripe state.
From this, it can be deduced that the pressure reduces the stripe feature toward the inhomogeneous charge and spin structure in the high pressure limit.
This is consistent with the physical picture of the HSD-LSZ transition discussed above.

We now make an estimate of the transition point within a meanfield approximation.
The hopping energy at the transition point corresponds to a lattice parameter change of about 8\% ($\frac{a}{a'}=\lambda^\frac{1}{3.5}\approx1.08$).
From the Birch-Murnaghan equation of state~\cite{birch} and the bulk modulus of NiO~\cite{huang}, the corresponding pressure is estimated as about 70-90 GPa.
Although the estimated transition point is quite high, the actual value should be lower, since the estimate is within a meanfield approximation.

%\section{summary}
In summary, we have investigated pressure effects on striped nickelate compounds by considering a four-band Hubbard model and allowing for a static displacement of oxygen ions.
Using a selfconsistent Hartree-Fock calculation, we found a pressure-induced transition from an HS stripe state to an LS one.
The LS stripe state is characteristic to nickelate system, where two $d$ orbitals are involved to make hole stripes, unlike those in cuprates, which involve only one $d$ orbital.

The prediction of this transition could be verified by observing a change in the x-ray diffraction pattern or in the phonon spectra.
The former is useful if the state changes into zigzag stripes.
If the stripe orientation does not change at the transition, the latter can provide signatures of the transition.
At the transition point, a softening of the lattice excitations around 60-65 $t_0$ meV ($\approx$ 484-524 $t_0$ cm$^{-1}$) by $7t_0$ meV ($\approx 56t_0$ cm$^{-1}$) should be observable by means of infrared measurements under high pressure.
Measurement by neutron scattering should also find a sharp change in the phonon spectra.
However, the transition point may be too high for a neutron experiment.

The phonon excitations in the LS stripe state have less stripe signature than in the HS stripe state.
We speculate that the LS stripe state may be metallic and that the transition has a metal-insulator transition character.
If so, the optical conductivity should show a change near the transition point.

%\section{acknowledgement}
We are grateful to Yang Ding for helping to estimate the critical pressure, and to Maiko Kofu, Masanori Ichioka, and Michel van Veenendaal for valuable discussions.
This work was supported by the U.S. DOE, Office of Science, Office of Basic energy Sciences, under Contract No. DE-AC02-06CH11357.

\end{document}